\documentstyle[12pt,epsf]{article}
\baselineskip=\normalbaselineskip
\def \Tr {{\ \rm Tr \ }}

\def \ln {{\rm \ ln \  }}

\def \ee {{\rm e}^}

\def \k1 {{1\overk}}

\def \ep {\epsilon}

\def \k {\kappa }

\def \m{\mu}
\def \n{\nu}

\def \e#1 {{\rm e}^{#1}}

\def \1p {{1\over  \pi }}
\def \2p {{{1\over  2\pi }}}
\def \4p {{ {1\over 4 \pi }}}
\def \8p {{{1\over 8 \pi }}}

\ifx\TwoupWrites\UnDeFiNeD\else\target{\magstepminus1}{11.3in}{8.27in}
	\source{\magstep0}{7.5in}{11.69in}\fi
\newfont{\fourteencp}{cmcsc10 scaled\magstep2}
\newfont{\titlefont}{cmbx10 scaled\magstep3}
\newfont{\authorfont}{cmcsc10 scaled\magstep1}
\newfont{\fourteenmib}{cmmib10 scaled\magstep2}
	\skewchar\fourteenmib='177
\newfont{\elevenmib}{cmmib10 scaled\magstephalf}
	\skewchar\elevenmib='177
\makeatletter
\newcommand\nonsequentialeqnum{
	\@addtoreset{equation}{section}
	\def\theequation{\arabic{section}.\arabic{equation}}}
\newif\ifp@bblock  \p@bblocktrue
\newcommand\nopubblock{\p@bblockfalse}
\newcommand\topspace{\hrule height 0pt depth 0pt \vskip}
\newcommand\p@bblock{\begingroup \tabskip=\hsize minus \hsize
	\baselineskip=1.5\ht\strutbox \topspace-2\baselineskip
	\halign to\hsize{\strut ##\hfil\tabskip=0pt\crcr
	\the\Pubnum\crcr\the\date\crcr}\endgroup}
\newcommand{\frontpageskip}{\vspace{12pt plus .5fil minus 2pt}}
\renewcommand{\title}[1]{\frontpageskip
	\begin{center}{\titlefont #1}\end{center}\par}
\renewcommand{\author}[1]{\frontpageskip\par\begin{center}
	{\authorfont #1}\end{center}
	\nobreak
	}

\newcommand{\address}[1]{\par\begin{center}{\sl #1}\end{center}\par}

\renewcommand{\thanks}[1]{\footnote{#1}}
\renewcommand{\abstract}{\par\frontpageskip\centerline{\fourteencp Abstract}
	\vspace{8pt plus 3pt minus 3pt}}
\begin{document}
\titlepage
\begin{tabbing}
\` FIT-HE-95-32 \\
\` KYUSAN-TH-381 \\
\` August 1995\\
\` hep-th/xxxxxxxx
\end{tabbing}

\renewcommand{\thefootnote}{\fnsymbol{footnote}}
\title{
On the validity of ADM formulation \\
 in 2d quantum gravity\\
}

\author{
K.\ Ghoroku${}^{1\,}$\thanks{
e-mail address: gouroku@dontaku.fit.ac.jp},
K.\ Kaneko${}^{2\,}$\thanks{
e-mail address: kaneko@daisy.te.kyusan-u.ac.jp}
}

\address{
${}^1$
Fukuoka Institute of Technology\\
Wajiro, Higashi-ku, Fukuoka 811-02, Japan \\
${}^2$
Kyushu Sangyo University\\
Matsukadai, Fukuoka 813, Japan\\
}

\renewcommand{\thefootnote}{\arabic{footnote}}
\setcounter{footnote}{0}
\newcommand{\cleqn}{\setcounter{equation}{0} \indent}
\renewcommand{\theequation}{\thesection.\arabic{equation}}
\newcommand{\beqa}{\begin{eqnarray}}
\newcommand{\eeqa}{\end{eqnarray}}
\newcommand{\nn}{\nonumber \\ }
\newcommand{\eq}[1]{(\ref{#1})}
\newcommand{\cD}{{\cal D}}
\newcommand{\cH}{{\cal H}}
\newcommand{\Psid}{\Psi^\dagger}
\newcommand{\norm}[1]{{\parallel {#1} \parallel}^2}
\newcommand{\nnorm}[1]{{{\parallel {#1} \parallel}^{\prime\,2}_l}}
\newcommand{\del}{\partial}
\newcommand{\db}{{\bar{\delta}}}
\newcommand{\gbar}{{\bar{g}}}
\newcommand{\dl}
           {\left[\,\frac{dl}{\,l\,}\,\right]}
\newcommand{\Det}{\,\mbox{Det}\,}
\newcommand{\ldot}{\dot{l}}
\newcommand{\bra}[1]{\left\langle\,{#1}\,\right|}
\newcommand{\ket}[1]{\left|\,{#1}\,\right\rangle}

\begin{abstract}

We investigate 2d gravity quantized in the ADM formulation, where only the
loop length $l(z)$ is retained as a dynamical variable
of the gravitation, in order to get an intuitive physical insight
of the theory.
The effective action of $l(z)$ is calculated by adding scalar fields
of conformal coupling, and the problems of the critical dimension
and the time development of $l$ are addressed.

\end{abstract}

\newpage
\section{Introduction}
\cleqn

Two-dimensional (2d) quantum gravity has been extensively studied
for the last few years, since it serves as a toy model of 4d quantum
gravity as well as a prototype of string theories.
In spite of great progress, there remains a difficulty in
2d quantum gravity.
There appears a strong restriction on matters coupled to
gravity \cite{kpz} \cite{ddk}.
In fact, until now we have no way to couple the matters whose central
charge is greater than one. In order to find a breakthrough in this
direction, various kinds of trials would be necessary.

There have been mainly two continuum approaches so far.
One is based on the conformal gauge \cite{ddk}, and the other
on the light-cone gauge \cite{kpz}.
Recently an approach, which could provide a more intuitive
understanding of the dynamics, based on
the ADM (Arnowitt-Deser-Misner) formalism has been proposed
\cite{nino} \cite{kawa},
and the time development of the loop variable was discussed.

In the present paper, we reconsider this formalism in terms of a
slightly different but simple gauge condition, where only the loop variable
is retained as the dynamical variable. In this formalism,
the time direction is treated specially so that we can easily
see the time development of the loop. But the lack
of the covariance is fatal since it is difficult to
get the quantum measure of the diffeomorphism
with a definite form. In fact, a non-renormalizable divergent term
appears if we proceed a straightforward calculation of the measure
without any care of the properties of the
operators. One way to get a meaningful and
definite answer is to impose a consistent restriction on the operators.
Such an example of a self-consistent calculation is
given here. This example gives the same result with \cite{nino},
where the induced action has a divergent pre-factor, so it leads
to a functional delta function, {\it i.e.\,}
a strong constraint on the loop variable. And this constraint is
equivalent to the former restriction imposed on the operators.
\par

On the other hand, the matter fields give a finite
induced action. As a result, the resultant effective
action is dominated by the quantum measure of the
diffeomorphism. And the scalar fields could not give any
effect on the induced action. This seems to be strange
since the effective action is insensitive to the number of the scalar
fields.
So it could be said that the formalism based on the ADM
decomposition is not suitable for
seeing the dynamical effect of the matter fields
on the surface. Therefore, the critical number of the scalar fields
(the critical dimension in terms of the string theory terminology),
which is obtained in the conformal gauge \cite{p},
can not be determined in this formalism.
\section{Quantum measure and ADM decomposition}
\cleqn
Here we firstly consider the case of pure gravity.
ADM decomposition is the most popular and convenient one in general relativity,
and it is given as follows in terms of
a metric $g_{\mu\nu}$ (with
Euclidian signature) on a two-dimensional manifold,
\beqa
	ds^2&=&g_{\mu\nu}\,dx^\mu dx^\nu\nn
	    &=&(Ndx^0)^2\,+\,h\,(\lambda dx^0+dx^1)^2, \label{2.1}
\eeqa
{\it i.e.\,},
\beqa
 	g_{\mu\nu}~=~
        \left[\begin{array}{cc}
	N^2+h\lambda^2 & h\lambda \\
        h\lambda & h
	\end{array}\right]\, . \label{2.2}
\eeqa
Here $N$ and $\lambda$ are the lapse- and shift-functions,
and $h$ is the metric on the time slice at $x^0$.

According to ref.\cite{nino}, the amplitude of a cylinder
with two loop boundaries $C$ and $C'$ of their length $l_0$ and $l_1$
respectively is defined as follows,
\beqa
   F(l_1,l_0) \equiv \int
   \frac{\cD g_{\mu\nu}}{{\rm Vol}_{diff}}\,
   \exp\left\{-\mu_0\int d^2x\sqrt{g}\right\} \, , \label{2.3}
\eeqa
where $\mu_0$ denotes the bare cosmological constant, and the constraints
\[  \delta\left(\int_C\sqrt{g_{\mu\nu}dx^\mu
   dx^\nu}-l_0\right)\,
   \delta\left(\int_{C'}\sqrt{g_{\mu\nu}dx^\mu dx^\nu}-l_1\right)
  \]
are abbreviated since they are not necessary hereafter. The functional measure
\[
  \cD g_{\mu\nu}=\cD N\cD h \cD \lambda
\]
is defined by the following norm,
\beqa
   \norm{\delta g_{\mu\nu}}_g~&\equiv~&\int d^2x\,\sqrt{g}\,
   g^{\mu\alpha}g^{\nu\beta}\,\delta g_{\mu\nu}|_g\,\delta
   g_{\alpha\beta}|_g\, \nn
     &~=~& 4\,\int d^2x\,N\sqrt{h}\,\left[\,
   \left(\frac{\delta h}{2h}\right)^2\,+\,
   \left(\frac{\delta N}{N}\right)^2\,+\,
   \frac{1}{2}\,\left(\frac{\sqrt{h}}{N}\,\delta\lambda\right)^2
   \,\right]\,.  \label{2.4}
\eeqa
The volume of diffeomorphism group ${\rm Vol}_{diff}$ is identified with
the functional integral over all diffeomorphisms,
\beqa
   {\rm Vol}_{diff}~\equiv~\int\cD V_\mu\,, \label{2.5}
\eeqa
where $\cD V_\mu$ is determined by the norm of infinitesimal
diffeomorphism $(x^\mu\,\mapsto\,x^\mu-\delta V^\mu(x))$ with
\beqa
   \norm{\delta V_\mu }_g~=~\int d^2x\,\sqrt{g}\,g^{\mu\nu}\,
   \delta V_\mu \delta V_\nu.
                \label{2.6}
\eeqa

Next, we take the following gauge fixing condition, $\gbar_{\mu\nu}$,
\beqa
   \bar{N}&\equiv&1\,, \label{2.7} \\
   \bar{\lambda}&\equiv&k={\rm constant}\,, \label{2.8}
\eeqa
or equivalently,
\beqa
   \gbar_{\mu\nu}(x^0=t,x^1=x)~\equiv~
   \left[\begin{array}{cc}
    1+l(t,x)^2k^2 & l(t,x)^2k \\
    l(t,x)^2k & l(t,x)^2
   \end{array}\right]. \label{2.9}
\eeqa
Here only $l(t,x)^2$ ($=\,\bar{h}$)
is retained as a free variable since $k$ is taken as a constant.
In \cite{nino}, $k$ is retained as a variable and $l$ is restricted as
$l=l(t)$. This point is essencially different from our gauge condition.
Eq.\ \eq{2.7} implies that
the time coordinate $x^0$ is chosen
directly as the geodesic distance from the incoming loop $C$.
In this gauge, the variations of the original three
independent variables, $\left\{N,\lambda,h\right\}$, are
replaced by the two gauge parameters, $\delta V_\mu$, and $\delta l$.

In the following, we take the coordinate
$(x^0,x^1)=(t,x)$ on $M$, such that $0\leq x\leq1$, and the times at
$C$ and $C'$
are represented by $t=0$ and $t=D$, respectively.
For this parametrization, it may be useful to introduce two
vectors $\vec{n}$ and $\vec{s}$ which are normal and tangential to
time slices, respectively;
\beqa
   n_\mu~\equiv~(1,0),~~~s_\mu~\equiv~l(k,1). \label{2.10}
\eeqa
Then eq.\ \eq{2.6} is rewritten at
$g_{\mu\nu}=\gbar_{\mu\nu}$ in the following form:
\beqa
   \norm{\delta V_\mu}_\gbar~=~\int d^2x\,l\,
   \left[ (\delta v^n)^2+(\delta v^s)^2\,\right]\,, \label{2.11}
\eeqa
where $\delta v^n$ and $\delta v^s$ are infinitesimal diffeomorphisms in the
normal and tangential directions, respectively,
\beqa
   \delta v^n~\equiv~\gbar_{\mu\nu}\,n^\mu\,\delta V^\nu\,,~~~
   \delta v^s~\equiv~\gbar_{\mu\nu}\,s^\mu\,\delta V^\nu\,.\label{2.12}
\eeqa
Eq.\ \eq{2.11} implies that ${\rm Vol}_{diff}$ depends on $l$, but its $l$
dependent part is expected as,
\[  \exp\left\{\alpha\delta(0)\int d^2x\ln(l)\right\} , \]
where $\alpha=-1/2$. And this divergent term
can be neglected if it is regularized by the
dimensional reguralization since it can be set zero. As a result,
${\rm Vol}_{diff}$ can be considered as a $l$-independent infinite number,
so it can be factored out and be devided out as in the usual gauge theory.

Further, since the infinitesimal deformation of metric around
$\gbar_{\mu\nu}$ is generally expressed as\footnote{
$\overline{\nabla}_\mu$ is the covariant derivative with respect to
$\gbar_{\mu\nu}$.
}
\beqa
   \delta g_{\mu\nu}~=~\delta\gbar_{\mu\nu}\,+\,
   \overline{\nabla}_\mu\delta V_\nu\,+\,
   \overline{\nabla}_\nu\delta V_\mu, \label{2.13}
\eeqa
we obtain the following relations by using eqs.\ \eq{2.2} and \eq{2.9},
\beqa
   \delta h&=&2l(\delta l-\Omega v^n+lD_sv^s), \nn
   \delta \lambda &=& {1 \over l}[(D_n+\Omega)v^s+D_sv^n], \nn
    \delta N&=&D_nv^n .  \label{2.14}
\eeqa
Here, $\Omega = -{1 \over l}D_nl$, and $D_n$, $D_s$ are
the derivatives in the normal and tangential
directions defined as,
\beqa
   D_n&\equiv&n^\mu\del_\mu~=~\del_0\,-\,k\,\del_1, \nn
   D_s&\equiv&s^\mu\del_\mu~=~l^{-1}\,\del_1\,. \label{2.16}
\eeqa
Their hermite conjugates, $D_n^\dagger$ and $D_s^\dagger$, under
the diffeomorphism-invariant measure\footnote{
We use the following abbreviation:
\beqa
   \dot{f}~\equiv~\del_0 f,~~~f'~\equiv~\del_1 f. \n
\eeqa
}~
$\int d^2x\,\sqrt{\gbar}\,=\,\int d^2x\,l$ are given as follows:
\beqa
   D_n^\dagger&=&-\,\del_0\,+\,k\,\del_1\,-\,\frac{\ldot}{l}\,,\nn
   D_s^\dagger&=&-\,l^{-1}\,\del_1\,.
   \label{2.17}
\eeqa
Then the original three components of $\delta g_{\m\n}$ can be
replaced by $\delta l$ and two diffeomorphism generators. This change of
variables is given by
\beqa
    \cD h\cD \lambda \cD N=\cD l\cD v^n\cD v^s
       \biggl|\frac{\partial(h, \lambda, N)}{\partial(l, v^s, v^s)}\biggr|,
     \  \      \label{2.18}
\eeqa
where the Jacobian can be written as,
\beqa
       \biggl|\frac{\partial(h, \lambda, N)}{\partial(l, v^s, v^s)}\biggr|
       =\Det^{1/2}D_n^\dagger D_n \cdot
        \Det^{1/2}(D_n^\dagger +\Omega)(D_n+\Omega). \label{2.19}
\eeqa
Here a numerical constant is abbreviated.
We thus obtain
\beqa
   \frac{\cD g_{\mu\nu}}{{\rm Vol}_{diff}}
    =\cD l \cdot\Det^{1/2}D_n^\dagger D_n \cdot
   \Det^{1/2}(D_n^\dagger +\Omega)(D_n+\Omega). \label{F8}
\eeqa
\section{Evaluation of the measure and the amplitude}
\cleqn

Next, our task is to evaluate the following functional of $l$:
\beqa
   F[l_1, l_0]~=~\int
   \cD l\cdot\Det^{1/2}D_n^\dagger D_n \cdot
   \Det^{1/2}(D_n^\dagger +\Omega)(D_n+\Omega)
    e^{-\m_0\int d^2zl}.  \label{E1}
\eeqa
The operators in the determinant are lacking the covariance, but
we can estimate them from the invariant operators
in terms of a limiting procedure adopted in \cite{nino}
and the method in \cite{p2}.
We can rewrite
the Laplacian in the manifold defined by $\gbar_{\m\n}$ as,
\beqa
   \Delta[\gbar_{\mu\nu}]&\equiv&-\frac{1}{\sqrt{\gbar}}\del_\mu
   \left(\sqrt{\gbar}\gbar^{\mu\nu}\del_\nu\right) \nn
   &=&D_n^\dagger D_n\,+\,D_s^\dagger D_s. \label{3.1}
\eeqa
Noticing that $\Delta[\gbar_{\mu\nu}]=\Delta[l,k]$, and rescaling $l$
by a positive constant $\beta$, we find the following relation,
\[
   \Delta[\beta^{-1}l,k]~=~D_n^\dagger D_n\,+\,
   \beta^2D_s^\dagger D_s\, .
\]
Then $D_n^\dagger D_n$ can be obtained from the Laplacian as,
\beqa
   D_n^\dagger D_n[l,k]~\equiv~\lim_{\beta\rightarrow+0}\,
   \Delta[\beta^{-1}l,k]\,. \label{3.2}
\eeqa
So $\Det^{1/2}D_n^\dagger D_n$ can be derived from the determinant
of the Laplacian (3.2) through the the above relation.
As for the operator $(D_n^\dagger +\Omega)(D_n+\Omega)$, it seems
difficult to get a similar relation as above. However,
the following relation can be easily found if we take $k=0$;
\beqa
   (D_n^\dagger +\Omega)(D_n+\Omega)
   [l,k]~\equiv~\lim_{\beta\rightarrow+0}\,
   \bar{\Delta}[\beta^{-1}l,k] \, , \label{G1}
\eeqa
where
\beqa
   \bar{\Delta}[g_{\mu\nu}]~\equiv~
         \Delta[g_{\mu\nu}]-{1 \over 2}R
    +g^{\mu\nu}\partial_{\mu}(\ln\sqrt{g})\partial_{\nu}(\ln\sqrt{g})
      \,. \label{G2}
\eeqa
So we evaluate (3.1) restricting to the case of $k=0$, but
the essential part of the theory will not be changed even if we
take $k\neq 0$. Then
we consider hereafter the case of $k=0$ for the sake of the brevity.

First, we estimate the determinant of the Laplacian
\[
   \Delta\,[g_{\mu\nu}]~\equiv~-\frac{1}{\sqrt{g}}
   \del_\mu(\sqrt{g}g^{\mu\nu}\del_\nu) .
\]
It is defined by
\beqa
   \ln \Det \Delta\,[g_{\mu\nu}]~&\equiv&~-\,\int_{\epsilon}^\infty
   \frac{ds}{s}\,\Tr\,e^{-s\Delta\,[g_{\mu\nu}]}, \label{B.2}
\eeqa
where $\epsilon$ is an urtraviolet cutoff. Its infinitesimal
change under the Weyl
transformation $g_{\mu\nu}\mapsto e^{2\delta\sigma}g_{\mu\nu}$
can be derived as follows \cite{alva},
\beqa
   \delta\,\ln\Det\Delta\,[g_{\mu\nu}]
   &\equiv&-2{\rm Tr}(\delta \sigma e^{-\epsilon\Delta}) \nn
   &=&-\,2\int d^2x\,\sqrt{g}\,\delta\sigma\,\left(
   \frac{1}{4\pi\epsilon}\,+\,\frac{1}{24\pi}R\,[g_{\mu\nu}]\,+\,
   O(\epsilon)\right)\,. \label{B.3}
\eeqa
Making use of this formula, we can obtain
the result as a functional of $l$ if the infinitesimal
deformation $\delta l$ is related to the
infinitesimal Weyl transformation $\delta\sigma$. For this purpose,
consider the following reparametrization \cite{nino},
$x^\mu\mapsto\tilde{x}^\mu(x)$, with
\beqa
   ds^2&=&e^{2\delta\sigma(t)}\gbar_{\mu\nu}dx^\mu dx^\nu \nn
   &=&e^{2\delta\sigma(t)}\,\left[
   (dt)^2\,+\,l(t)^2\left(k(t,x)dt+dx\right)^2\right] \nn
   &\equiv&(d\tilde{t})^2\,+\,\tilde{l}^2\left(
   \tilde{k}d\tilde{t}+d\tilde{x}\right)^2\,.\label{B.4}
\eeqa
Since $k$ is not a variable but
a constant \footnote{
Here we retain $k$ as non-zero as far as possible, but it is
set to be zero at some stage as shown below.}
, then we take $\tilde{k}=k$. Then the possible parametrization
which satisfies the above requirement is obtained as follows,
\[
      \tilde{x}(t,x)\equiv x,
\]
with
\beqa
   {\del\tilde{t} \over \del t}&=&1+(1-l^2k^2)\delta\sigma\,, \nn
   \tilde{l}&=&l[1+(1+l^2k^2)\delta\sigma]\, , \label{B.6}
\eeqa
where the terms of $O(\delta\sigma^2)$ are neglected.
Thus we obtain
\beqa
   \delta l&\equiv&\tilde{l}(x,t)\,-\,l(x,t) \nn
   &=&l(1+l^2k^2)\delta\sigma-{\del l \over \del t}\delta\tau ,
   \label{B.7}
\eeqa
where
\beqa
   \delta\tau&\equiv&\tilde{t}\,-\,t\nn
   &=&\int_0^tdt'\,(1-l^2k^2)\delta\sigma(t',x)\,.\label{B.8}
\eeqa

Here we take $k=0$, which is necessary for the calculation of
$ {\bar \Delta}\,[g_{\mu\nu}] $.
But the physical consequences would not depends on the value of
$k$, so this is a shortcut to get a meaningful result.
Thus we obtain from the above equations
the following expression for $\delta\tau(t,x)$ and $\delta\sigma(t,x)$,
\beqa
   \delta\tau(t,x)=l(t,x)\,\int_0^tdt'\,
           \frac{\delta l(t,x')}{l(t,x')^2}\,,
   \label{B.9}
\eeqa
\beqa
   \delta\sigma(t,x)={\del l(t,x) \over \del t}
       \int_0^tdt'\,\frac{\delta l(t,x')}{l(t,x')^2}\,+\,
   \frac{\delta l(t,x)}{l(t,x)}\,. \label{B.10}
\eeqa
Furthermore, the scalar curvature $R$ is
given by
\beqa
   R\,[\gbar_{\mu\nu}]~=~-\,2\,\frac{\ddot{l}}{\,l\,}\, \,.\label{B.11}
\eeqa
for $g_{\m\n}=\gbar_{\m\n}$ and $k=0$.
Finally, we arrive at the following result by using
the eqs.\eq{B.3}, \eq{B.10} and \eq{B.11},
\beqa
   \delta\,\ln\Det\Delta\,[\gbar_{\mu\nu}]~=~-\,
   \delta\int dz\,\left[\,\mu_1\,l(z)\,+\,
   \frac{1}{12\pi}\,\frac{\ldot(z)^2}{l(z)}\,\right]\,.
   \label{B.15}
\eeqa
where $\mu_1\equiv \lim_{\epsilon \rightarrow 0}1/4\pi\epsilon$ is the
divergent term.

Next, we evaluate the determinant of ${\bar \Delta}(g_{\m\n})$ in a
similar way to the case of $\Delta(g_{\m\n})$.
However we must notice that this operator is not positive
definite for arbitrary values of $g_{\m\n}$ due to the second term of
\eq{G2}
even if the negative modes of $\Delta(g_{\m\n})$ are excluded. So we must
assume the positive definiteness
of the operator ${\bar \Delta}(g_{\m\n})$ and
examine
the validity of this assumption after the calculation.

If we assume that the eigenvalue of ${\bar \Delta}$ is positive,
then the following formula is
obtained,
\beqa
   \delta\,\ln\Det{\bar\Delta}\,[g_{\mu\nu}]
   &=&-2{\rm Tr}(\delta \sigma e^{-\epsilon{\bar\Delta}}) \nn
     & & \qquad +{\rm Tr}({\bar\Delta}^{-1}e^{-\epsilon{\bar\Delta}}
         [\nabla^2+4g^{\m\n}\del_{\mu}\ln\sqrt{g}\del_{\nu}]\delta\sigma)
                           \,. \label{G.3}
\eeqa
The first term of \eq{G.3}
is written as,
\beqa
   -2{\rm Tr}(\delta \sigma e^{-\epsilon{\bar\Delta}})
   =\int\, d^2z\sqrt{g(z)}\delta\sigma(z)
              G(z,z';\epsilon)\, \, ,\label{G.4}
\eeqa
where
\beqa
   G(z,z';\epsilon)
   \equiv <z|e^{-\epsilon{\bar\Delta}}|z'>\, \, . \label{G.5}
\eeqa
We estimate this diffusion evolution operator $G(z,z';\epsilon)$
according to the method of \cite{alva}
by taking the conformal flat metric, $g_{\mu\nu}=e^{2\sigma}\delta_{\m\n}$,
since the results are rewritten in a general form.
In conformal coordinates,
\beqa
     {\bar\Delta}&=&-\del^2-{\bar V}          \nn
     {\bar V}&=& (e^{-2\sigma}-1)\del^2
                -e^{-2\sigma}[\del^2\sigma+4(\del\sigma)^2]\, \, ,
         \label{G.6}
\eeqa
where the second term of ${\bar V}$ is the newly appeared one compared
to the case of the operator $\Delta$.
Further, choose a
local conformal coordinate with the origin being the intersting point,
\beqa
    \sigma(0)&=&0 \,\, ,   \nn
    \del_{\m}\sigma(0)&=&0 \, \, . \label{G.7}
\eeqa
Then we obtain the following result after a straightforward calculation,
\beqa
   G(0,0;\epsilon)={1 \over 4\pi\epsilon}+{1 \over 6\pi}R
                     +O(\epsilon) \, .  \label{G.8}
\eeqa
Comparing \eq{G.8} and \eq{B.3}, we can see the difference of the
coefficient of $R$.

As for the second term of \eq{G.3}, we write it as follows,
\beqa
   \int\, d^2z\sqrt{g(z)}\delta\sigma(z)
         [\nabla^2-4g^{\m\n}\del_{\mu}\ln\sqrt{g}\del_{\nu}]
           <z|{\bar\Delta}^{-1}e^{-\epsilon{\bar\Delta}}|z>
                           \, , \label{G.9}
\eeqa
by making use of partial integrations. In the calculation, we expand
the inverse operator ${\bar\Delta}$ as,
\[ {\bar\Delta}={1 \over -\del^2}
       +{1 \over -\del^2}{\bar V}{1 \over -\del^2}+\cdots .\]
After a straightforward calculation, we get
\beqa
   &&{\rm Tr}({\bar\Delta}^{-1}e^{-\epsilon{\bar\Delta}}
         [\nabla^2+4g^{\m\n}\del_{\mu}\ln\sqrt{g}\del_{\nu}]\delta\sigma) \nn
   &&     = \int\, d^2z\sqrt{g(z)}\delta\sigma(z)
         [{-1 \over 16\pi}(2+\Gamma(0)\ln(0^+))R+O(\epsilon)]
                           \, , \label{G.10}
\eeqa
where the divergent coefficient $\Gamma(0)$ appeared because of the
infrared divergence of the loop of "massless propagator" $1/\del^2$ in
${\bar \Delta}^{-1}$ expansion. So this could be regularized
by introducing some small cutoff parameter. However the divergence due to
$\ln(0^+)$ is urtraviolet, and this divergence can not be absorbed by the
cosmological constant since this term is corresponding
to the anomaly term. This result seems to indicate the
non-renormalizability of the theory. But this
statement would not be correct, and we should consider
that our assumption of the positive definiteness of \eq{G2}
is broken at some point. Then \eq{G.3} would not be correct.

In order to proceed the estimation of the measure, we restrict the last
two terms of \eq{G2} such that they can be written as
\beqa
   \bar{\Delta}[g_{\mu\nu}]=
         \Delta[g_{\mu\nu}]+m^2
      \,. \label{F2}
\eeqa
This replacement of the operator $\bar{\Delta}[g_{\mu\nu}]$ is corresponding
to giving a restriction on the loop variable as,
\beqa
  l(x,t)=l_0(x){\rm exp}(\pm{m \over \sqrt{2}}t)\, . \label{F3}
\eeqa
It can be seen that this restriction is consistent with the final result
of the calculation of the determinants as seen below. And it should be noted
that the restriction \eq{F3} is implied by the calculation of
$\Det^{1/2}D_n^\dagger D_n$. If we could perform the calculation of
$\Det^{1/2}(D_n^\dagger+\Omega)(D_n+\Omega)$ under a more loose restriction,
it might lead to an interesting result. But it is beyond our present work.

In this case, the eq.\eq{G.3} is changed as follows,
\beqa
   \delta\,\ln\Det{\bar\Delta}\,[g_{\mu\nu}]
   &=&-2{\rm Tr}(\delta \sigma e^{-\epsilon{\bar\Delta}}) \nn
     & & \qquad +2m^2{\rm Tr}({\bar\Delta}^{-1}e^{-\epsilon{\bar\Delta}}
                             \delta\sigma)
                           \,. \label{F4}
\eeqa
Then we obtain after a straightforward calculation the following form,
\beqa
   \delta\,\ln\Det{\bar\Delta}\,[g_{\mu\nu}]
    &=&-2 \int\, d^2z\sqrt{g(z)}\delta\sigma(z)
     \biggl({1 \over 4\pi\epsilon}
        +{m^2 \over 4\pi}[1+\ee{m^2\ep}{\rm E}_i(-m^2\ep)] \nn
       & & \qquad +{1 \over 24\pi}R
                     +O(\epsilon)\biggr) \, ,  \label{F5}
\eeqa
where ${\rm E}_i$ is the error function,
\[{\rm E}_i(-x)=-\int_x^{\infty}{dt \over t}\ee{-t} .\]
And \eq{F5} is easily rewritten in terms of the variable $l$,
\beqa
   \delta\,\ln\Det{\bar\Delta}\,[\gbar_{\mu\nu}]~=~-\,
   \delta\int d^2z\,\left[\,\mu_2\,l(z)\, \,
   +\frac{1}{12\pi}\,\frac{\ldot(z)^2}{l(z)}\,\right]\, ,
   \label{G.12}
\eeqa
where
\beqa
   \m_2={1 \over 4\pi\epsilon}
        +{m^2 \over 4\pi}[1+\ee{m^2\ep}{\rm E}_i(-m^2\ep)]
                         \, .  \label{G.13}
\eeqa

{}From \eq{B.15} and \eq{G.12}, we get the following final result,
\beqa
   F[\,l\,]&=&\int \cD l e^{-S_{\rm eff}(l)}  \nn
   S_{\rm eff}(l)&=&{c \over 24\pi}
         \int d^2z\,\biggl[\frac{\ldot(z)^2}{l(z)}
           -{m^2 \over 2}l(z)\biggr]\,
         +\m^2\int d^2z\,l(z)\,    ,\label{G.14}
\eeqa
where the constant $c$ is defined as,
\beqa
   c~\equiv~\lim_{\beta,\gamma\rightarrow+0}\,
   ({1 \over \beta}+{1 \over \gamma})\, , \label{G.15}
\eeqa
and $\mu^2$ is the renormalized cosmological constant.
The parameter $m$ has been taken into the resultant formula because of
the consistency of the previous assumption \eq{F3}
which was adopted in the calculation
of Det${\bar \Delta}$. It can be easily seen that
we can obtain the relation \eq{F3} from \eq{G.14} since
$c$ is divergent. Then we arrive at
\beqa
   S_{\rm eff}(l)=\m^2\int d^2z\,l(z)\,    ,\label{G.16}
\eeqa
where the constraint \eq{F3} was used.
This implies the area law of the time
development amplitude of the loop, and the loop develops according
to \eq{F3}. It indicates an inflation or a deflation
of one dimensional universe depending on the sign in \eq{F3}.

\section{Matter fields and the critical dimension}
\cleqn

Here we consider the cilinder amplitude $F[l]$
by including the matter fields contribution. The action is written
as follows in this case,
\beqa
  S=\m_0^2\int d^2z\sqrt{g}
      +\int d^2z\sqrt{g}{1 \over 2}g^{\mu\nu}
           \del_{\m}\phi_i\del_{\n}\phi_i\, , \label{4.1}
\eeqa
where $i=1\sim d$. The measure of the scalars is defined by the norm,
\beqa
   \norm{\delta \phi_i}_g~=~\int d^2x\,\sqrt{g}\,
             (\delta \phi_i)^2\, .
                \label{4.2}
\eeqa
Then the term,
$-{d \over 2}\ln\Det{\Delta}\,[g_{\mu\nu}]$, is added to the effective
action after integration over the scalar fields. So, $S_{\rm eff}$ in \eq{G.14}
is modified here as follows,
\beqa
      S_{\rm eff}(l)={c-d \over 24\pi}
         \int d^2z\,\biggl[\frac{\ldot(z)^2}{l(z)}
           -{m^2 \over 2}l(z)\biggr]\,
                 +\m^2\int d^2z\,l(z)\,    ,\label{4.3}
\eeqa
where the same notation is used for the renormalized cosmological
constant $\m^2$, and $c$ is given in \eq{G.15}.

As shown in the previous section, the value of $c$ is infinite. So we
arrive at the same result \eq{F3} and \eq{G.16} for any value of $d$.
Then the scalar field could not give any influence on the effective action,
but this conclusion seems to be contradicted with the analyses performed
in terms of the conformal gauge where the coefficient of the induced
Liouville term is proportional to $26-d$. The reason why we could not
get this critical dimension would be reduced to the incomplete calculation of
Det${\bar \Delta}$. It has been obtained under a
consistent constraint on $l(z)$
and the result has the same form with the one of Det$\Delta$. However
there is a possibility that we obtain a finite $c$ if it has the form,
\beqa
   c~\equiv~\lim_{\beta,\gamma\rightarrow+0}\,
   ({1 \over \beta}-{c_1 \over \gamma})\, , \label{G15}
\eeqa
where $c_1$ is some positive number. In this case, both the infinities
might cancel out leaving a finite number. And the critical dimension
could be determined if $c$ is obtained as a definite number, which is
expected to be 26.

{}From the result obtained so far, the classical path of $l$ is constrained
by \eq{F3}. In order to investigate the quantum mechanical problem of
the loop dynamics, the quantum measure of $l$ must be determined.
However it seems to be difficult
since the norm of $\delta l$ is defined as
\beqa
   \norm{\delta l}_g~=~\int d^2x\,l(\delta l)^2\, ,
                \label{4.9}
\eeqa
then the measure has $l$-dependence.
In the case of the conformal gauge, this problem was solved by exploiting
conformal invariance with respect to the fiducial metric. But there
is no such a principle to determine the measure here. So we could not
proceed the work to find the so-called
dressed factor, which has been given in the case of the conformal gauge
formulation, of the perturbation operator like
the cosmological term. As shown in the case of the
conformal gauge [5], it is also expected that a new instability of the
surface would be seen if we could calculate the string susceptibility.
These analyses seem to be impossible here.

\section{Conclusion and Discussions}
\cleqn

We have examined the formulation of 2d gravity
in terms of the ADM decomposition in order
to see the time development of the cylinder, the closed one dimensional
space. This gauge is unfortunately
not covariant, so it is difficult to see the
critical dimension, the critical number of the scalar fields
which couple to the gravity
with the conformal coupling. In fact, we can not determine the coefficient of
the induced kinetic term of the loop variable, $l$, coming from
the measure of the diffeomorphism. The main reason of this difficulty
is in the lack of the covariance of the quadratic operators.
In our self-consistent calculation, the coefficient of the kinetic term
of $l$ in the induced action is divergent.
This situation means that
this coefficient is
not affected by the quantum effect of scalar fields
even if we add any number of scalar fields. This result implies that
the formalism based on ADM decomposition is not valid to see the dynamics
of the surface which should be sensible to the fields on the surface.

\newpage

\newpage

\end{document}